\documentclass[9pt,conference]{IEEEtran}
\IEEEoverridecommandlockouts

\usepackage{cite}
\usepackage{amsmath,amssymb,amsfonts}
\usepackage{algorithmic}
\usepackage{graphicx}
\usepackage{textcomp}
\usepackage{xcolor}
\def\BibTeX{{\rm B\kern-.05em{\sc i\kern-.025em b}\kern-.08em
    T\kern-.1667em\lower.7ex\hbox{E}\kern-.125emX}}

\usepackage{color}
\usepackage{amsmath}
\usepackage{comment}
\usepackage{multirow}
\usepackage{amsmath}
\usepackage{amsfonts}
\usepackage{colortbl}
\usepackage{cite}
\usepackage{arydshln}
\usepackage{url}
\usepackage{mathrsfs}
\usepackage{subfigure}
\usepackage{subfig}
\usepackage{color}

\usepackage{amssymb}
\usepackage{amsfonts}
\usepackage{mathtools}
\usepackage{graphics}

\newcommand{\ten}[1]{ \boldsymbol{\mathcal #1}}

\newlength\savedwidth
\newcommand{\wcline}[1]{\noalign{\global\savedwidth\arrayrulewidth\global\arrayrulewidth 1.0pt} \cline{#1}
\noalign{\global\arrayrulewidth\savedwidth}}
\usepackage{graphics}

\begin{document}

\makeatletter
\newcommand{\linebreakand}{%
  \end{@IEEEauthorhalign}
  \hfill\mbox{}\par
  \mbox{}\hfill\begin{@IEEEauthorhalign}
}
\makeatother

\title{Trainingless Adaptation of Pretrained Models\\for Environmental Sound Classification}

\author{

\IEEEauthorblockN{1\textsuperscript{st} Noriyuki Tonami}
\IEEEauthorblockA{
\textit{NEC Corporation}\\
Japan
}
\and
\IEEEauthorblockN{2\textsuperscript{nd} Wataru Kohno}
\IEEEauthorblockA{
\textit{NEC Laboratories America, Inc.}\\
The United States of America
}
\and
\IEEEauthorblockN{3\textsuperscript{rd} Keisuke Imoto}
\IEEEauthorblockA{
\textit{Doshisha University}\\
Japan
}
\linebreakand
\IEEEauthorblockN{4\textsuperscript{th} Yoshiyuki Yajima}
\IEEEauthorblockA{
\textit{NEC Corporation}\\
Japan
}
\and
\IEEEauthorblockN{5\textsuperscript{th} Sakiko Mishima}
\IEEEauthorblockA{
\textit{NEC Corporation}\\
Japan
}
\and
\IEEEauthorblockN{6\textsuperscript{th} Reishi Kondo}
\IEEEauthorblockA{
\textit{NEC Corporation}\\
Japan
}
\and
\IEEEauthorblockN{7\textsuperscript{th} Tomoyuki Hino}
\IEEEauthorblockA{
\textit{NEC Corporation}\\
Japan
}

}

\maketitle

\begin{abstract}
Deep neural network (DNN)-based models for environmental sound classification are not robust against a domain to which training data do not belong, that is, out-of-distribution or unseen data.
To utilize pretrained models for the unseen domain, adaptation methods, such as finetuning and transfer learning, are used with rich computing resources, e.g., the graphical processing unit (GPU).
However, it is becoming more difficult to keep up with research trends for those who have poor computing resources because state-of-the-art models are becoming computationally resource-intensive.
In this paper, we propose a trainingless adaptation method for pretrained models for environmental sound classification.
To introduce the trainingless adaptation method, we first propose an operation of recovering time--frequency-ish (TF-ish) structures in intermediate layers of DNN models.
We then propose the trainingless frequency filtering method for domain adaptation, which is not a gradient-based optimization widely used.
The experiments conducted using the ESC-50 dataset show that the proposed adaptation method improves the classification accuracy by 20.40 percentage points compared with the conventional method.
\end{abstract}


\section{Introduction}
The analysis of environmental sounds has been studied for various applications \cite{Imoto_AST2018_01,Mesaros_SPmaga2021_01}.
Analyzing environmental sounds enables machine condition monitoring \cite{Dohi_dcase2023}, traffic monitoring \cite{damiano_icassp2024}, and bioacoustic analysis \cite{Nolasco_dcase2023}. 

Deep neural network (DNN)-based models for analyzing environmental sounds have been attracting attention for the past ten years \cite{SED_CRNN,kong_TASLP2020_01,miyazaki_icassp2020_01,gong_interspeech_2021}.
In \cite{SED_CRNN}, convolutional neural network bidirectional gated recurrent unit (CNN-BiGRU) has been used for sound event detection (SED).
In \cite{kong_TASLP2020_01,miyazaki_icassp2020_01}, Transformer is employed to handle longer-sequence audio.
In \cite{gong_interspeech_2021}, Vision Transformer (ViT) \cite{Dosovitskiy_iclr2021} has resulted in a significant improvement in environmental sound classification.

DNN-based models are not robust against unseen data, which are the data that the models have not encountered during the training stages. 
To address robustness against unseen data, adaptation methods, such as finetuning and transfer learning, have been studied. 
In the analysis of environmental sounds, domain adaptation for environmental sound classification \cite{Lopez_icassp2021}, acoustic scene classification (ASC) \cite{Heittola_dcase2020,He_access2021,Martin_dcase2022,Tan_taslp2024}, and the detection of anomalous machine conditions \cite{Dohi_dcase2023} has been in progress.
In \cite{Heittola_dcase2020}, ASC for addressing the gap in recording devices between training and inference stages has been studied.
In \cite{He_access2021}, the generative adversarial network (GAN)-based domain adaptation for optical fiber sensing with ASC has been introduced.
In the detection of anomalous machine conditions, domain generalization has been developed in the DCACE2023 challenge task 2 \cite{Dohi_dcase2023}.

More recently, low-calculation-cost adaptation methods have been studied \cite{Martin_dcase2022,hu_iclr2022} for tackling the problem that the scale of a dataset and DNN models are becoming larger.
In the DCASE2022 challenge task 1 \cite{Martin_dcase2022}, the participants of the task develop low-calculation-cost adaptation methods with gradient-based optimization.
Low-rank adaptation (LoRA) \cite{hu_iclr2022,Zheng_arxiv} has been proposed for addressing the finetuning of large-scale models.
However, it is assumed that the adaptation methods including LoRA require rich computing resources, e.g., a graphical processing unit (GPU).
In such a situation, only organizations with rich computing resources can follow research trends and realize rapid developments.

\begin{figure}[t!]
  \centering
  \includegraphics[width=.499\textwidth]{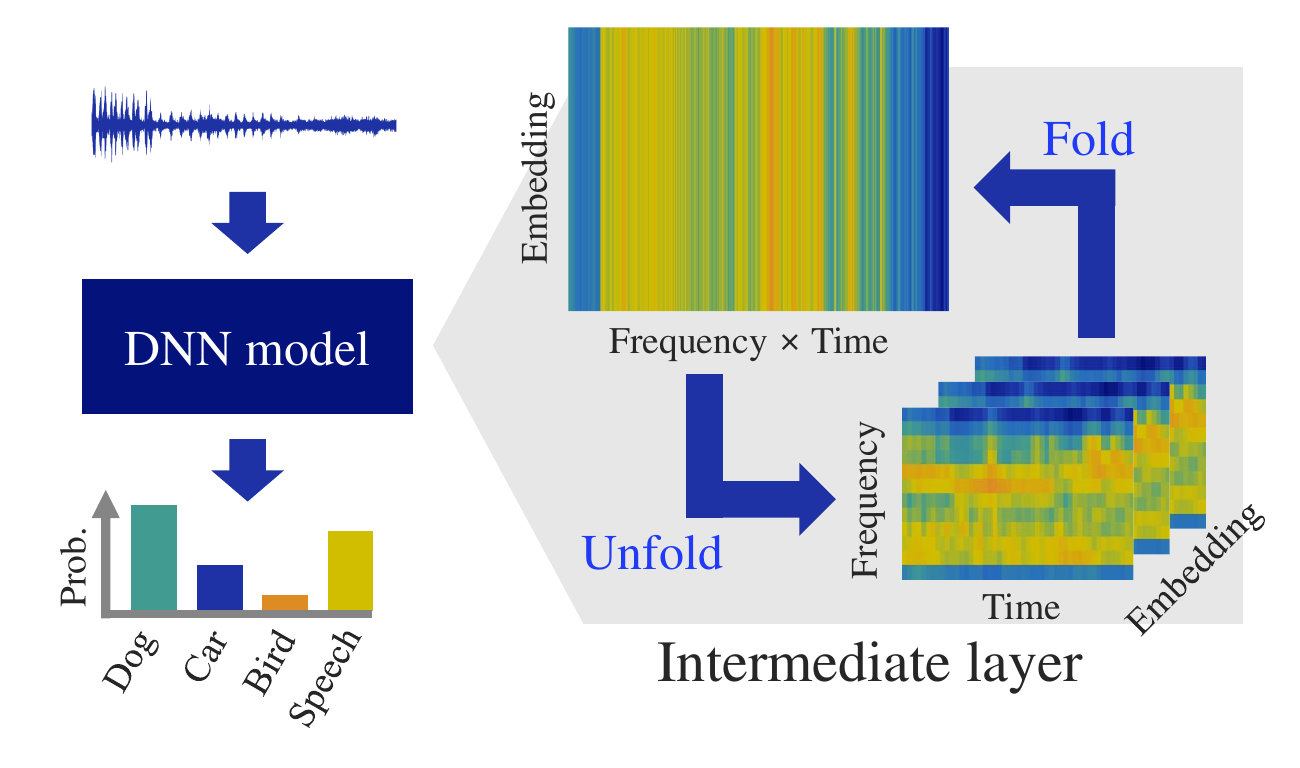}\\
  \vspace{-8pt}
  \caption{Concept of TF-ish structures in intermediate layer of DNN}
  \vspace{-5pt}
  \label{fig:concept}
\end{figure}

In this paper, we propose a trainingless adaptation method for pretrained models for environmental sound classification.
For the implementation of the trainingless adaptation method, we first propose a method of recovering time--frequency-ish (TF-ish) structures in intermediate layers of DNN models.
The trainingless frequency filtering method for domain adaptation is then proposed, as shown in Fig. \ref{fig:concept}.
By the frequency filtering of the signal processing, which is not a gradient-based optimization, DNN-based models acquire robustness against unseen data without high calculation cost requiring training data.
In this paper, we did not use any GPUs for experiments.

\begin{figure*}[t!]
  \centering
  \includegraphics[width=0.94\textwidth]{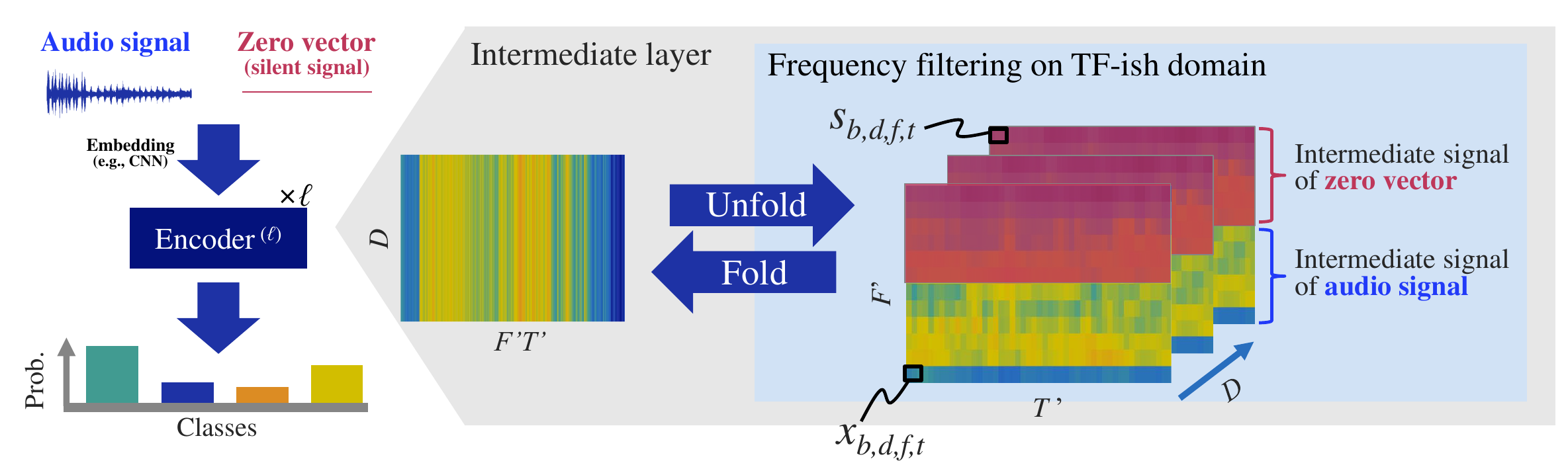}\\
  \vspace{-5pt} 
  \caption{Proposed frequency filtering method on TF-ish domain of intermediate layers}
  \vspace{-5pt}
  \label{fig:detail_proposed_filtering}
\end{figure*}

\vspace{0pt}
\section{Conventional method}
\vspace{0pt}

\subsection{Audio spectrogram transformer (AST)}
Recently, in the task of environmental sound classification, AST models, which are based on ViT \cite{Dosovitskiy_iclr2021}, have been widely used \cite{gong_interspeech_2021,koutini_interpseech2022,Chen_icml2023,Kolesnikov_icassp2024}.
The output of the $\ell$-th layer in AST $\mathcal{X}^{(\ell)} \in \mathbb{R}^{B \times D \times F^{\prime} T^{\prime}}$ is formalized as
\begin{align}
\vspace{-3pt}
    \label{eq:layer}
    \mathcal{X}^{(\ell)} = \textsf{Encoder}^{(\ell)}(\mathcal{X}^{(\ell-1)}).
\vspace{-3pt}
\end{align}
\noindent
$B$ and $D$ represent the batch size and the number of dimensions of the embedding, respectively.
$F^{\prime}$ and $T^{\prime}$ indicate the numbers of partitions of the patches on the frequency and time axes, respectively.
$\textsf{Encoder}^{(\ell)}$ is the nonlinear operator.
Note that Eq. \ref{eq:layer} is formulated without a special token, such as the class token.
In an AST-based architecture, it is mainly composed of multihead attention, time-distributed multi-layer perceptron (MLP), and layer normalization.
These functions of $\textsf{Encoder}^{(\ell)}$ save the shape of an inputted feature, which is a log-mel spectrogram in the time--frequency domain.
The AST models are trained using large-scale audio datasets, such as AudioSet \cite{Gemmeke_ICASSP2017}.
The AST models are expected to contain TF-ish features of intermediate layers.

\subsection{Adaptation of DNN models}
For DNN models, many adaptation methods, including finetuning, have been developed \cite{Heittola_dcase2020,He_access2021,Lopez_icassp2021,Martin_dcase2022,Tan_taslp2024,Zheng_arxiv}.
Finetuning is employed to mitigate the gap between the distributions of training and inference data, where the gap might cause the underperformance of classifying sound events.
The finetuning generally utilizes gradient-based optimization to adapt models to unseen data using a dataset.
These gradient-based adaptation methods require heavy calculations, which make it difficult to rapidly cycle the development for updating the parameters of the models.

\vspace{0pt}
\section{Proposed method}
\vspace{0pt}
We introduce the adaptation method with which the pretrained models for environmental sound classification acquire robustness against unseen data without gradient-based optimization.
The proposed adaptation method employs a naive signal processing in the intermediate layers focusing on the TF-ish structure appearing in AST.
Fig. \ref{fig:detail_proposed_filtering} shows the proposed trainingless adaptation method with which the frequency filtering is reproduced for extracting informative frequency bands in the intermediate layers.

\subsection{Recovering TF-ish structure in intermediate layer}
We first define two operations for handling TF-ish features in the intermediate layers of AST.
In AST models, spectrograms $\ten{X}$ of $F$ frequency bins and $T$ time frames are first embedded using the convolutional neural network (CNN): $\mathbb{R}^{B \times 1 \times F \times T} \rightarrow \mathbb{R}^{B \times D \times F \times T} $.
The embedded spectrograms are then reshaped with the following operation: 
\begin{align}
\vspace{0pt}
 \textsf{Fold} \coloneqq \mathbb{R}^{B \times D \times F^{\prime} \times T^{\prime}} \rightarrow \mathbb{R}^{B \times D \times F^{\prime}T^{\prime}}.
\vspace{0pt}
\end{align}
\noindent
The reshaped $\textsf{Fold}(\ten{X})$ is known as the ``patch.''
We further define the following operation in the intermediate layers of AST $\mathcal{X}^{(\ell)}$:
\begin{align}
\vspace{0pt}
 \textsf{Unfold} \coloneqq \mathbb{R}^{B \times D \times F^{\prime} T^{\prime}} \rightarrow \mathbb{R}^{B \times D \times F^{\prime} \times T^{\prime}}.
\vspace{0pt}
\end{align}
\noindent
This operation, where a third-order tensor is transformed into a fourth-order one, indicates the reverse of $\textsf{Fold}(\cdot)$.
In this work, we use a simple reshape operator as $\textsf{Fold}(\cdot)$, which is exactly the reverse operator of $\textsf{Fold}(\cdot)$ for the widely used AST models \cite{gong_interspeech_2021,koutini_interpseech2022,Chen_icml2023,Elizalde_icassp2022}.
By applying the $\textsf{Unfold}$ operation in the intermediate layers of AST, we can expect to recover TF-ish structures.

\subsection{Signal processing in intermediate layers with TF-ish structure}
To make the pretrained models robust against unseen data without retraining, we also introduce the signal-processing-based adaptation method.
In the pretrained DNN models, the intermediate signals are normalized, e.g., the layer normalization with pretrained parameters.
Such normalization could be regarded as a denoising operator, which enhances informative low-power signals.
However, non-informative signals are also enhanced together with informative low-power signals.
The enhanced non-informative signals then cause the misclassification of environmental sounds. 
In the proposed adaptation method, the non-informative signals are not enhanced by switching non-informative frequency bands into silent signals in intermediate layers on the TF-ish domain.

The adaptation method for the reshaped intermediate layers $\textsf{Unfold}(\mathcal{X}^{(\ell)})$ of AST for reproducing a naive signal processing is as follows:
\begin{align}
\vspace{0pt}
    \label{eq:prop}
    x^{(\ell)}_{b,d,f,t} \leftarrow s^{(\ell)}_{b,d,f,t}  \mid_{ {f=c} ~\in \{0,\dots, F^{\prime}-1\} },
\vspace{0pt}
\end{align}
\noindent
where $x^{(\ell)}_{b,d,f,t}$ and $s^{(\ell)}_{b,d,f,t} \in \mathbb{R}$ respectively represent the outputs of the intermediate layers for the inputted audio and silent signals in terms of the $b$-th batch, $d$-th dimensional embedding, $f$-th bin frequency, and $t$-th time frame.
The signal $x^{(\ell)}_{b,d,f,t}$ of the intermediate layer of the inputted audio signal is switched into the signal $s^{(\ell)}_{b,d,f,t}$ of that of the inputted silent signal, i.e., the signal contains only the direct current component.
Eq. \ref{eq:prop} reproduces the filtering on the TF domain in the intermediate layers of AST models.
After applying Eq. \ref{eq:prop} to the intermediate layers, $\textsf{Fold}$ is again applied.

\begin{figure}[t!]
  \centering
  \includegraphics[width=.46\textwidth]{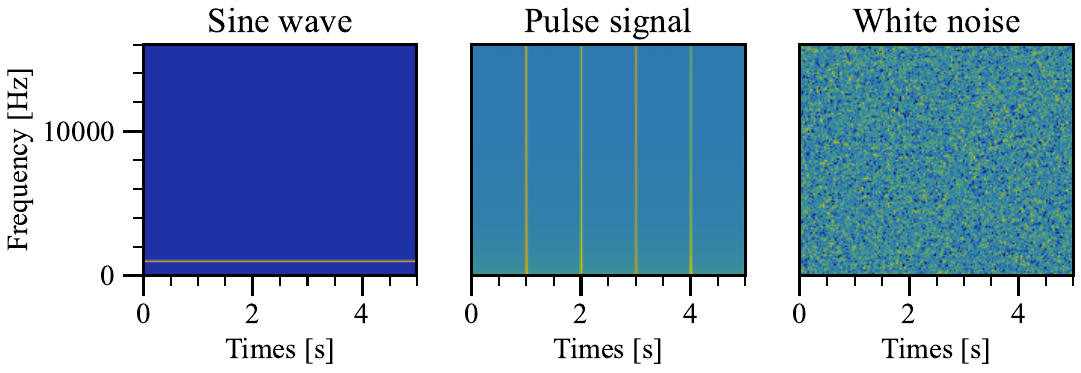}\\
  \caption{Signals used for confirmation of TF-ish features in intermediate layers}
  \vspace{0pt}
  \label{fig:signals}
\end{figure}

\begin{figure}[t!]
  \centering
  \includegraphics[width=.46\textwidth]{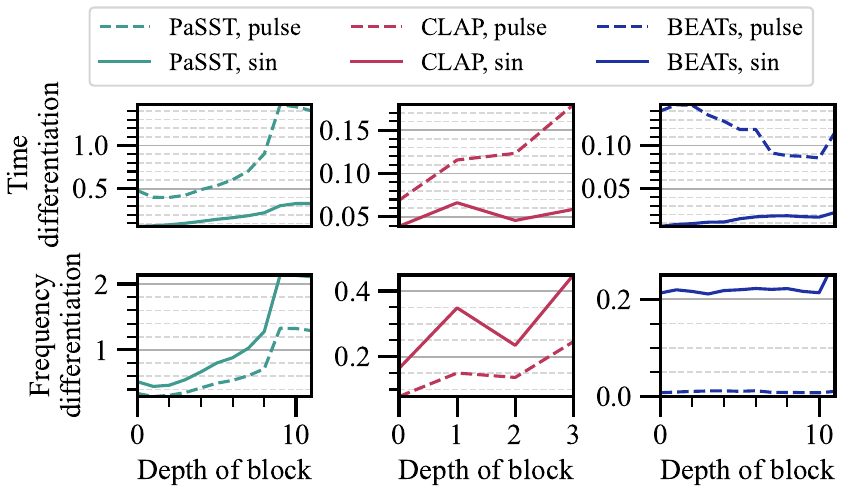}\\
  \caption{Time and frequency continuity in intermediate layers of AST models}
  \vspace{0pt}
  \label{fig:fig_TF_continuity}
\end{figure}

\begin{figure}[t!]
  \centering
  \includegraphics[width=.46\textwidth]{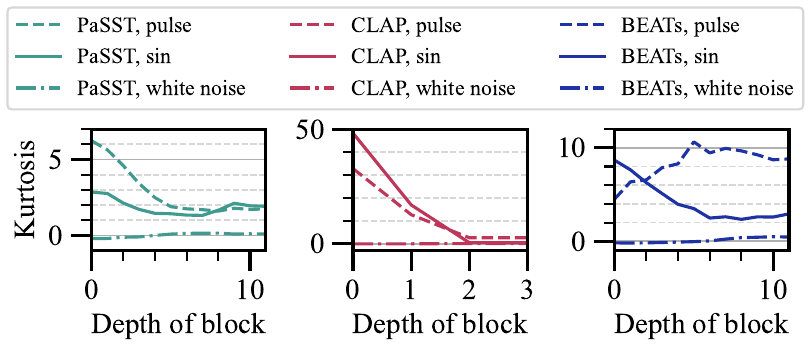}\\
  \caption{Randomness in intermediate layers of AST models}
  \vspace{3pt}
  \label{fig:fig_kurtosis}
\end{figure}

\begin{figure*}[t!]
  \centering
  \includegraphics[width=0.98\textwidth]{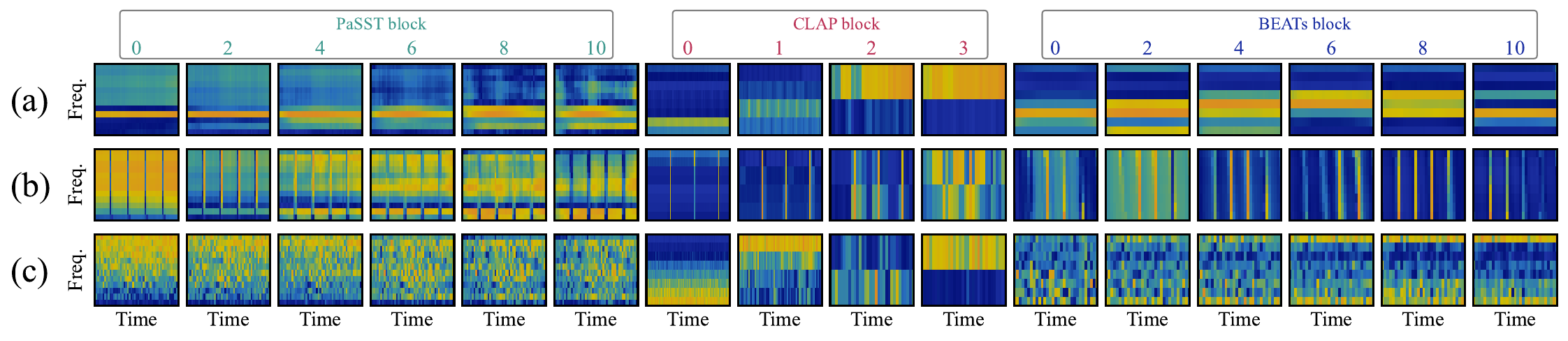}\\
  \vspace{-3pt} 
  \caption{Visualization of TF-ish features in intermediate layers}
  \vspace{-0pt}
  \label{fig:fig_visialization}
\end{figure*}

\vspace{0pt}
\section{Experiments}
\vspace{0pt}
\subsection{Experimental conditions}

\noindent
{\bf [Dataset, parameters of models, and training]}
In experiments, the proposed filtering of Eq. \ref{eq:prop} is used for classifying environmental sounds.
To evaluate the performance, we used ESC-50 \cite{Piczak_ACMM2015} with five-fold cross-validation.
ESC-50 is composed of 2,000 audio signals with 50 environmental sound classes.
For widely used baselines of AST models, we used Patchout fast Spectrogram Transformer (PaSST) \cite{koutini_interpseech2022}, Contrastive language-audio pretraining  \cite{Kolesnikov_icassp2024}, referred to as ``MS-CLAP,'' and Bidirectional encoder representation from audio Transformers (BEATs) \cite{Chen_icml2023}.
PaSST, MS-CLAP, and BEATs are supervised, audio-language supervised, and self-supervised models, respectively.
For PaSST, we used the pretrained weights of ESC-50, which are officially provided.
For MS-CLAP, the events of ESC-50 are classified by prompts, e.g., ``{\it this is a sound of [class label]},'' in accordance with  \cite{Elizalde_icassp2022} using hierarchical token semantic audio Transformer (HTS-AT) \cite{Chen_icassp2022} and generative pretrained Transformer 2 (GPT2) \cite{radford2019language} as the audio and text encoders, respectively.
For BEATs, we finetuned the model initialized with the parameters of ``BEATs\_iter3+ (AS2M)'' \cite{Chen_icml2023}. 
The AST models used in the experiments did not conduct data augmentation by adding noise to the training dataset.

\noindent
{\bf [Target domain]}
In the experiments, we consider the Distributed Fiber-Optic Sensor (DFOS) \cite{Lu_ICASSP2021,tonami_dcase2023,tonami_icassp2024} as the target domain, which is different from that of the microphone.
In DFOS, acoustic signals are captured by an optical fiber.
Regarding the characteristics of DFOS, the observed audio signals are from low-pass filters covered by cables and optical noise with low signal-to-noise Ratio (SNR) \cite{tonami_dcase2023}.
The quality of audio signals observed by DFOS is thus significantly lower than that of audio signals observed by the microphone.

For reproducing audio signals observed by DFOS, we applied 1st- to 4th-order Butterworth low-pass filters to audio signals of ESC-50.
To low-frequency-pass audio signals, white Gaussian noises with variable SNRs are added to reproduce optical noise.
The larger the order of the filters and the lower the SNR, the more different the domain of DFOS is from that of the microphone.

\noindent
{\bf [Setting for proposed method]}
In Eq. \ref{eq:prop}, the frequency bin $f$ is calculated as
\begin{align}
\vspace{0pt}
  m = \underset{m} {\operatorname{argmin}} (f_c - f_m) \\
  f^{\prime}_{c}  = \lfloor F^{\prime}m / M \rfloor,
\vspace{0pt}
\end{align}
\noindent
where $f_c$ and $f_m$ indicate the cutoff frequency of the Butterworth low-pass filter used for DFOS and the center frequency of mel-frequency bins, which are used for the preprocessing of the AST models.
$f^{\prime}_{c}$ and $M$ represent the index of the cutoff frequency bins of the intermediate layers and the number of mel-frequency bins, respectively.

Moreover, we use the term ``block'' in the experiment to refer to the intermediate layer for each AST model.
For PaSST and BEATs, one Transformer block corresponds to one ``block.''
For the HTS-AT of MS-CLAP, one group (Swin Transformer + Patch-Merge) corresponds to one ``block.''

\begin{table*}[t!]
  \vspace{2pt}
  \centering
  \caption{Summary results of environmental sound classification in terms of accuracy [\%]}
  \vspace{0pt}
  \label{tbl:result_summary}
  \scalebox{0.99}{
  \begin{tabular}{cccccccccccccccc}
  \wcline{1-16}
  &&&&&&&&&&&&&&& \\[-6pt]
  & & &&& &  & \multicolumn{4}{c}{SNR} & &&& &  \\
  & & \multicolumn{4}{c}{$-$5 dB} &  & \multicolumn{4}{c}{$-$10 dB} &  & \multicolumn{4}{c}{$-$15 dB} \\ \cline{3-6}\cline{8-11}\cline{13-16} 
  &&&&&&&&&&&&&&& \\[-8pt]
  & & \multicolumn{4}{c}{Order for low-pass filter} &  & \multicolumn{4}{c}{Order for low-pass filter} &  & \multicolumn{4}{c}{Order for low-pass filter} \\ 
  & \multirow{-4}{*}{Model} & 1 & 2 & \multicolumn{1}{c}{3} & \multicolumn{1}{c}{4} &  & 1 & \multicolumn{1}{c}{2} & \multicolumn{1}{c}{3} & \multicolumn{1}{c}{4} &  & 1 & \multicolumn{1}{c}{2} & \multicolumn{1}{c}{3} & \multicolumn{1}{c}{4} \\ \hline
  &&\cellcolor[HTML]{C0C0C0}&\cellcolor[HTML]{C0C0C0}&\cellcolor[HTML]{C0C0C0}&\cellcolor[HTML]{C0C0C0}&\cellcolor[HTML]{C0C0C0}&\cellcolor[HTML]{C0C0C0}&\cellcolor[HTML]{C0C0C0}&\cellcolor[HTML]{C0C0C0}&\cellcolor[HTML]{C0C0C0}&\cellcolor[HTML]{C0C0C0}&\cellcolor[HTML]{C0C0C0}&\cellcolor[HTML]{C0C0C0}&\cellcolor[HTML]{C0C0C0}&\cellcolor[HTML]{C0C0C0} \\[-8pt]
  Conventional & & \multicolumn{1}{c}{\cellcolor[HTML]{C0C0C0}68.35} & \multicolumn{1}{c}{\cellcolor[HTML]{C0C0C0}66.45} & \cellcolor[HTML]{C0C0C0}65.55 & \cellcolor[HTML]{C0C0C0}65.05 & \cellcolor[HTML]{C0C0C0} & \multicolumn{1}{c}{\cellcolor[HTML]{C0C0C0}49.55} & \cellcolor[HTML]{C0C0C0}48.20 & \cellcolor[HTML]{C0C0C0}48.00 & \cellcolor[HTML]{C0C0C0}47.70 & \cellcolor[HTML]{C0C0C0} & \multicolumn{1}{c}{\cellcolor[HTML]{C0C0C0}26.40} & \cellcolor[HTML]{C0C0C0}26.05 & \cellcolor[HTML]{C0C0C0}25.70 & \cellcolor[HTML]{C0C0C0}25.35 \\ 
  Proposed & \multirow{-2}{*}{PaSST} & \multicolumn{1}{c}{\cellcolor[HTML]{EFEFEF}76.05} & \cellcolor[HTML]{EFEFEF}75.60 & \cellcolor[HTML]{EFEFEF}75.05 & \cellcolor[HTML]{EFEFEF}74.25 & \cellcolor[HTML]{EFEFEF} & \multicolumn{1}{c}{\cellcolor[HTML]{EFEFEF}65.30} & \cellcolor[HTML]{EFEFEF}65.65 & \cellcolor[HTML]{EFEFEF}65.45 & \cellcolor[HTML]{EFEFEF}65.05 & \cellcolor[HTML]{EFEFEF} & \multicolumn{1}{c}{\cellcolor[HTML]{EFEFEF}\bf44.60} & \cellcolor[HTML]{EFEFEF}\bf45.55 & \cellcolor[HTML]{EFEFEF}\bf45.95 & \cellcolor[HTML]{EFEFEF}\bf45.75 
	\\ \hline
  &&\cellcolor[HTML]{C0C0C0}&\cellcolor[HTML]{C0C0C0}&\cellcolor[HTML]{C0C0C0}&\cellcolor[HTML]{C0C0C0}&\cellcolor[HTML]{C0C0C0}&\cellcolor[HTML]{C0C0C0}&\cellcolor[HTML]{C0C0C0}&\cellcolor[HTML]{C0C0C0}&\cellcolor[HTML]{C0C0C0}&\cellcolor[HTML]{C0C0C0}&\cellcolor[HTML]{C0C0C0}&\cellcolor[HTML]{C0C0C0}&\cellcolor[HTML]{C0C0C0}&\cellcolor[HTML]{C0C0C0} \\[-8pt]
  Conventional & & \multicolumn{1}{c}{\cellcolor[HTML]{C0C0C0}60.15} & \multicolumn{1}{c}{\cellcolor[HTML]{C0C0C0}57.05} & \cellcolor[HTML]{C0C0C0}55.55 & \cellcolor[HTML]{C0C0C0}54.70 & \cellcolor[HTML]{C0C0C0} & \multicolumn{1}{c}{\cellcolor[HTML]{C0C0C0}46.60} & \cellcolor[HTML]{C0C0C0}44.25 & \cellcolor[HTML]{C0C0C0}42.75 & \cellcolor[HTML]{C0C0C0}42.30 & \cellcolor[HTML]{C0C0C0} & \multicolumn{1}{c}{\cellcolor[HTML]{C0C0C0}30.15} & \cellcolor[HTML]{C0C0C0}28.65 & \cellcolor[HTML]{C0C0C0}28.15 & \cellcolor[HTML]{C0C0C0}27.85 \\ 
  Proposed & \multirow{-2}{*}{MS-CLAP} & \multicolumn{1}{c}{\cellcolor[HTML]{EFEFEF}59.95} & \cellcolor[HTML]{EFEFEF}58.25 & \cellcolor[HTML]{EFEFEF}57.45 & \cellcolor[HTML]{EFEFEF}56.80 & \cellcolor[HTML]{EFEFEF} & \multicolumn{1}{c}{\cellcolor[HTML]{EFEFEF}48.70} & \cellcolor[HTML]{EFEFEF}48.45 & \cellcolor[HTML]{EFEFEF}47.70 & \cellcolor[HTML]{EFEFEF}47.10 & \cellcolor[HTML]{EFEFEF} & \multicolumn{1}{c}{\cellcolor[HTML]{EFEFEF}\bf 36.20} & \cellcolor[HTML]{EFEFEF}\bf 36.25 & \cellcolor[HTML]{EFEFEF}\bf 36.75 & \cellcolor[HTML]{EFEFEF}\bf 36.70 
	\\ \hline
  &&\cellcolor[HTML]{C0C0C0}&\cellcolor[HTML]{C0C0C0}&\cellcolor[HTML]{C0C0C0}&\cellcolor[HTML]{C0C0C0}&\cellcolor[HTML]{C0C0C0}&\cellcolor[HTML]{C0C0C0}&\cellcolor[HTML]{C0C0C0}&\cellcolor[HTML]{C0C0C0}&\cellcolor[HTML]{C0C0C0}&\cellcolor[HTML]{C0C0C0}&\cellcolor[HTML]{C0C0C0}&\cellcolor[HTML]{C0C0C0}&\cellcolor[HTML]{C0C0C0}&\cellcolor[HTML]{C0C0C0} \\[-8pt]
  Conventional & & \multicolumn{1}{c}{\cellcolor[HTML]{C0C0C0}31.45} & \multicolumn{1}{c}{\cellcolor[HTML]{C0C0C0}29.80} & \cellcolor[HTML]{C0C0C0}28.90 & \cellcolor[HTML]{C0C0C0}28.65 & \cellcolor[HTML]{C0C0C0} & \multicolumn{1}{c}{\cellcolor[HTML]{C0C0C0}15.05} & \cellcolor[HTML]{C0C0C0}14.85 & \cellcolor[HTML]{C0C0C0}14.75 & \cellcolor[HTML]{C0C0C0}14.60 & \cellcolor[HTML]{C0C0C0} & \multicolumn{1}{c}{\cellcolor[HTML]{C0C0C0}5.85} & \cellcolor[HTML]{C0C0C0}5.70 & \cellcolor[HTML]{C0C0C0}5.90 & \cellcolor[HTML]{C0C0C0}5.75 \\ 
  Proposed & \multirow{-2}{*}{BEATs} & \multicolumn{1}{c}{\cellcolor[HTML]{EFEFEF}46.40} & \cellcolor[HTML]{EFEFEF}47.80 & \cellcolor[HTML]{EFEFEF}48.25 & \cellcolor[HTML]{EFEFEF}48.60 & \cellcolor[HTML]{EFEFEF} & \multicolumn{1}{c}{\cellcolor[HTML]{EFEFEF}\bf 33.50} & \cellcolor[HTML]{EFEFEF}\bf 34.75 & \cellcolor[HTML]{EFEFEF}\bf 35.55 & \cellcolor[HTML]{EFEFEF}\bf 35.70 & \cellcolor[HTML]{EFEFEF} & \multicolumn{1}{c}{\cellcolor[HTML]{EFEFEF}15.70} & \cellcolor[HTML]{EFEFEF}16.50 & \cellcolor[HTML]{EFEFEF}16.85 & \cellcolor[HTML]{EFEFEF}17.20 \\ \wcline{1-16}
\end{tabular}
  }
  \vspace{0pt}
\end{table*}

\vspace{0pt}
\subsection{Experimental results}
\vspace{0pt}

\noindent
{\bf [Quantitative confirmation of TF-ish features in intermediate layers]}
We first verify the assumption that TF-ish features are expressed in the intermediate layers of AST models.
To verify the assumption, we conduct experiments in terms of time continuity, frequency continuity, and randomness in the intermediate features.
More specifically, we inputted a sine wave, a pulse signal, and a white Gaussian noise of 5-s duration to the AST models.
For the sine wave and pulse signal, we used a sine wave of 1 kHz and a square signal of 1 Hz for a clear analysis.
Fig. \ref{fig:signals} shows the amplitude spectrograms of the sine wave, pulse signal, and white Gaussian noise used in the experiment. 

Fig. \ref{fig:fig_TF_continuity} shows the time and frequency continuity in the intermediate layers of the AST models.
As for the evaluation metrics of the time and frequency continuities, we used the time and frequency differentiations $\sum_{b}\sum_{d}\sum_{f}\sum_{t} |x_{b,d,f,t}-x_{b,d,f,t+1}|/BDF'T'$ and $\sum_{b}\sum_{d}\sum_{f}\sum_{t} |x_{b,d,f,t}-x_{b,d,f+1,t}|/BDF'T'$, respectively. 
For each intermediate layer of each AST model, the time and frequency differentiations were calculated.
For all models, the time differentiation of the sine wave is smaller than that of the pulse signal.
In other words, in the intermediate layers of the AST models, the TF-ish feature in terms of the time continuity is saved.
Similarly, the frequency differentiation of the pulse signal is smaller than that of the sine wave.
The result indicates that the continuity along the frequency axis is preserved in the intermediate layers. 

Fig. \ref{fig:fig_kurtosis} shows the response in the randomness in the intermediate layers.
In this experiment, white Gaussian noise was inputted to each AST model for the extra analysis.
Kurtosis was used to measure the sharpness of a distribution.
The kurtosis close to zero indicates that a signal follows the white Gaussian noise.
The kurtosis in Fig. \ref{fig:fig_kurtosis} was calculated by averaging the kurtosis of each frequency bin in each intermediate layer.
As shown in the figure, the kurtosis to the white Gaussian noise is closer to zero than those of the sine wave and pulse signal for each model and layer.
The result indicates that the intermediate layers of the AST models maintain the randomness of the TF structure.
Moreover, in the HTS-AT of MS-CLAP, the kurtosises of the pulse signal, sine wave, and white noise are close to each other.
This is because the number of dimensions on the TF-ish domain is reduced to decrease the calculation cost through its network architecture.

\noindent
{\bf [Qualitative confirmation of TF-ish features in intermediate layers]}
Fig. \ref{fig:fig_visialization} shows examples of the TF-ish features in the intermediate layers of the AST models.
For PaSST, MS-CLAP, and BEATs, $d=66, 18$, and $20$ of the intermediate layers are randomly selected and then shown in the figure.
Figs. \ref{fig:fig_visialization}(a) - \ref{fig:fig_visialization}(c) show the unfolded features in the intermediate layers for the inputted sine wave, pulse signal, and white Gaussian noise, respectively.
The results show that the features of the inputted signals are preserved in the intermediate layers.
In the deeper blocks, the features of the inputted signals tend to be destroyed.
Moreover, in the HTS-AT of MS-CLAP, it is clear that the number of dimensions on the TF-ish domain is reduced.

\noindent
{\bf [Summary results of event classification]}
We verify the proposed filtering of Eq. \ref{eq:prop} for classifying the environmental sounds hereafter.
Table \ref{tbl:result_summary} shows the summary results of the environmental sound classification in terms of accuracy.
In the table, ``conventional method'' and ``proposed method'' mean the AST model with and without the proposed filtering of Eq. \ref{eq:prop}, respectively.
The indexes of the ending block for the proposed filtering are 10, 2, and 3 for PaSST, MS-CLAP, and BEATs, respectively.
In the blocks shallower than the ending block, the proposed filtering is applied. 
As shown in the table, the proposed method outperforms the conventional method in most configurations. 
Under the condition of a lower SNR, the accuracy of the proposed method is significantly higher than that of the conventional method.
In particular, the proposed method improved the accuracy of PaSST by 20.40 percentage points compared with the conventional method at SNR$=-15$ dB and order$=$4. 
On the other hand, under the condition of SNR$=-5$ dB and order$=1,2$, the proposed filtering with MS-CLAP underperformed the conventional method.
This is because the audio encoder HTS-AT of MS-CLAP reduces the number of dimensions of the frequency in the deeper block. 
This reduction mechanism is helpful for decreasing the calculation cost.
However, the lower-dimensional frequency is harmful for frequency filtering.

\noindent
{\bf [Application of the proposed method to different layers]}
Fig. \ref{fig:fig_change_layers} shows the result of the change in classification performance by changing the ending block of the intermediate layers for the proposed filtering.
For example, in the left panel of the figure, the values at the 5th ending block represent the accuracies when the proposed filtering is applied from the 0th to 5th blocks in PaSST. 
In the results, only the average score of the orders 1 to 4 for the low-pass filter is discussed for a clear analysis. 
The results show that the accuracies of MS-CLAP and BEATs are significantly lower in the deeper ending blocks than in the shallower ending blocks.
This is because over-enhancing the inputted signals in the deeper blocks by MS-CLAP and BEATs, as shown in Fig. \ref{fig:fig_visialization}, leads to the destruction of the frequency structures.  

\begin{figure}[t!]
  \centering
  \includegraphics[width=.46\textwidth]{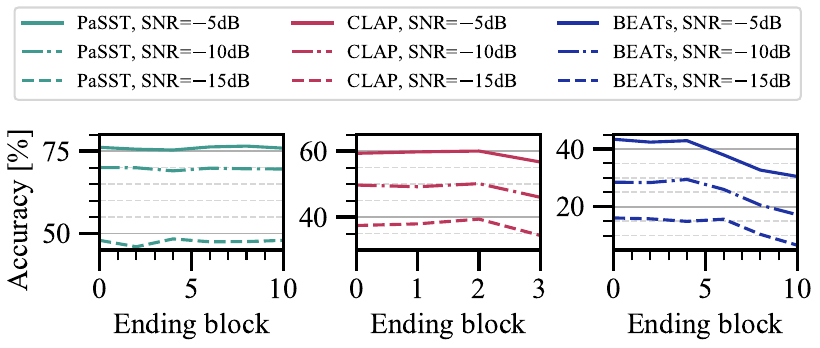}\\
  \caption{Change in classification performance by changing ending block of intermediate layers of proposed filtering}
  \vspace{-5pt}
  \label{fig:fig_change_layers}
\end{figure}

\vspace{0pt}
\section{Conclusion}
\vspace{0pt}
In this paper, we proposed the trainingless adaptation method for pretrained AST models.
In the experiments, we first confirmed the TF-ish features in the intermediate layers of AST.
We then proposed the frequency filtering method for the TF-ish domain to address the domain adaptation.
The results of experiments using the ESC-50 dataset show that the proposed filtering method improved the classification accuracy by 20.40 percentage points compared with the conventional method.

Our proposed method enables the application of signal processing methods to DNN models.
In other words, it accelerates the combination of legacy signal processing methods and modern DNN methods.

\clearpage
\bibliographystyle{IEEEbib}
\bibliography{IEEEabrv,refs_et_al}
\end{document}